\documentclass[11pt,a4paper,epsfig]{article}
\usepackage{color}
\usepackage{amsfonts}
\usepackage{epsfig}

\title{{\bf One-loop fluctuations}\\{\bf of} \\ {\bf semi-local self-dual vortices}}

\author{A. Alonso Izquierdo$^{(a)}$,
W. Garcia Fuertes$^{(b)}$\\ M. de la Torre Mayado$^{(c)}$, J. Mateos
Guilarte$^{(c)}$
\\ {\normalsize {\it $^{(a)}$ Departamento de Matematica
Aplicada and IUFFyM}, {\it Universidad de Salamanca,
SPAIN}}\\{\normalsize {\it $^{(b)}$ Departamento de Fisica} ,{\it
Universidad de Oviedo, SPAIN}}\\ {\normalsize {\it $^{(c)}$
Departamento de Fisica and IUFFyM} ,{\it Universidad de Salamanca,
SPAIN}}}
\date{}

\begin{document}
\maketitle
\begin{abstract}
Mass shifts induced by one-loop fluctuations of semi-local self-dual
vortices are computed. The procedure is based on canonical
quantization and heat kernel/zeta function regularization methods.
The issue of the survival of the classical degeneracy in the
semi-classical regime is explored.
\end{abstract}

\section{{\bf Introduction}}

In this communication we shall deal with one-loop mass shifts for
the semilocal self-dual topological solitons -SSTS in the sequel-
that arise in the (2+1)-dimensional semilocal Abelian Higgs model;
see \cite{VaAc1} for a review of the history and properties of these
classical solitonic backgrounds. On the analytical side, a formula
will be derived that involves the coefficients of the heat-kernel
expansion associated with the second-order fluctuation operator.
Additionally, numerical methods are used to generate the solutions
and to compute the coefficients. All this together will allow us to
obtain numerical results for one-loop SSTS mass shifts.

Control of the ultra-violet divergences arising in the procedure is
achieved by using heat kernel/zeta function regularization methods.
In the absence of detailed knowledge of the spectrum of the
differential operator governing second-order fluctuations around
vortices, the expansion of the associated heat kernel will be used
in a way akin to that developed in the computation of one-loop mass
shifts for one-dimensional kinks; see \cite{AJMW1}. In fact, a
similar technique has been applied previously to compute the mass
shift for the supersymmetric kink \cite{BRvNV}, although in this
latter case the boundary conditions must respect supersymmetry. In
the case of vortices, the only available results refer to either
supersymmetric vortices, achieved by Vassilevich and the Stony
Brook/Wien group, \cite{Vas}, \cite{RvNW}, or non-supersymmetric
self-dual vortices, obtained by our group, \cite{AJmW1}.

The closely related issue of computing the quantum energy of QED
flux tubes due to fermionic fluctuations has been addressed in
\cite{BD} and, more recently, in the papers \cite{W} and
\cite{GKQSW}. Quantum energies of the more subtle electroweak
strings caused by fermionic fluctuations have been thoroughly
studied in \cite{Weigel1} from a (2+1)-dimensional point of view for
$\theta=0$ Weinberg angle. We shall concentrate on the value
$\theta=\frac{\pi}{2}$. For this weak mixing angle the $SU(2)$ gauge
field decouples , the strings become topologically stable, and a
broader class of topological solitons arise because the Higgs vacuum
manifold becomes the $S^3$-sphere Hopf bundle. We shall restrict
ourselves, however, to consider only the bosonic fluctuations over
topological solitons saturating the Bogomolny bound. This is in
contrast to the work mentioned above where fermionic fluctuations
dominate because the fermions carry a high enough number of colors.

The study of quantum fluctuations of topological defects arising in
models that describe sub-atomic phenomena is a very important and
difficult subject. With the exceptions of sine-Gordon and
$\lambda(\phi)^4_2$ kinks, the knowledge of the spectrum of the
second-order differential operators governing these fluctuations is
non complete. Therefore, asymptotic methods, phase shifts,
high-temperature expansions, etcetera, must be used. In particular,
one must compute the $L^2$ trace of the square root of a
second-order differential operator, a problem for which the zeta
function/heat kernel regularization techniques, see \cite{EORBZ},
are specially suitable. Unfortunately, difficulties with this
procedure increase with the dimension of space-time. Nevertheless,
the experience with these planar examples makes conceivable the
possibility of computing the one-loop mass shift for BPS magnetic
monopoles sometime in the future.

\section{{\bf The planar semi-local Abelian Higgs model}}

We write the action governing the dynamics of the semi-local AHM in
the form{\footnote{Details of our conventions and calculations are
given in \cite{AJmW3}}}:
\[
S= \frac{v}{e}\int d^3 x \left[ -\frac{1}{4} F_{\mu \nu} F^{\mu
\nu}+\frac{1}{2} (D_\mu \Phi)^* D^\mu \Phi - \frac{\kappa^2}{8}
(\Phi^\dagger \Phi-1)^2 \right] \qquad .
\]
Besides the Abelian gauge field $A_\mu(x^\mu)$, there is a doublet
of complex scalar fields. The action is invariant with respect to
$U(1)$ gauge (local) and global (rigid) $SU(2)$ transformations, and
it is no more than the bosonic sector of the electro-weak theory
when the weak mixing angle is ${\pi\over 2}$. Note that we define
the electric charge unconventionally: $Q=-T_3+{1\over 2}Y$, in such
a way that the neutral scalar field is the upper component of the
weak iso-spinorial Higgs field.

A shift of the complex scalar field from the vacuum
\[
\Phi(x^\mu)=\left(\begin{array}{c}\Phi_1(x^\mu)\\
\Phi_2(x^\mu)\end{array}\right)=\left(\begin{array}{c}
\Phi_1^1(x^\mu)+i\Phi_1^2(x^\mu) \\
\Phi_2^1(x^\mu)+i\Phi_2^2(x^\mu)\end{array}\right)=\left(\begin{array}{c} 1+H(x^\mu)+iG(x^\mu) \\
\sqrt{2}\varphi(x^\mu)\end{array}\right)
\]
and choice of the Feynman-'t Hooft renormalizable gauge $R(A_\mu ,
G)=\partial_\mu A^\mu(x^\mu) - G(x^\mu)$ lead us to write the action
in terms of Higgs $H$, real Goldstone $G$, complex Goldstone
$\varphi$, vector boson $A_\mu$ and ghost $\chi$ fields:
\begin{eqnarray}
S+S_{{\rm g.f.}}+S_{{\rm ghost}}&=&{v\over e}\int \, d^3x \, \left[
-\frac{1}{2} A_\mu
[-g^{\mu\nu}(\Box +1)]A_\nu +\partial_\mu\chi^*\partial^\mu \chi- \chi^*\chi\right.\nonumber\\
&+&\frac{1}{2}\partial_\mu G\partial^\mu G-\frac{1}{2} G^2+
\frac{1}{2}\partial_\mu H\partial^\mu H-\frac{\kappa^2}{2}  H^2+\partial_\mu\varphi^* \partial^\mu \varphi\label{eq:act}\\
&-& {\kappa^2\over 2}H (H^2+G^2)+ A_\mu (\partial^\mu H
G-\partial^\mu G H)+H (A_\mu A^\mu -\chi^* \chi) +i A_\mu(\varphi^*
\partial^\mu \varphi-\varphi \partial^\mu \varphi^*)\nonumber\\&+& \left.
A_\mu A^\mu |\varphi|^2 -\frac{\kappa^2}{8} (H^2+G^2)^2
+\frac{1}{2}(G^2+H^2) A_\mu A^\mu
-\frac{\kappa^2}{2}|\varphi|^2(|\varphi|^2+H^2+G^2+2H)\right]\nonumber
\end{eqnarray}

\subsection{{\bf Vacuum energy}}

Canonical quantization promoting the coefficients of the plane wave
expansion around the vacuum of the fields to operators provides the
free quantum Hamiltonian. Besides the plane wave expansions in a
normalizing plate of very huge area $L^2$ of the fields of the
Abelian Higgs model considered in the third paper of Reference
\cite{AJmW1} we must also take into account the massless complex
Goldstone bosons:
\begin{itemize}

\item If $m=ev$,
\[
\delta \varphi(x_0,\vec{x})=\frac{e}{mL}\sqrt{\frac{\hbar}{m}}
\sum_{\vec{k}}\frac{1}{\sqrt{2\gamma(\vec{k})}}\left[f^*(\vec{k})e^{ikx}+g(\vec{k})e^{-ikx}\right]
\quad , \quad \gamma(\vec{k})=+\sqrt{\vec{k}\vec{k}}
\]
\[
[\hat{f}(\vec{k}),\hat{f}^\dagger(\vec{q})]=[\hat{g}(\vec{k}),\hat{g}^\dagger(\vec{q})]
=\delta_{\vec{k}\vec{q}}\Rightarrow
H^{(2)}[\delta\hat{\varphi}]=\hbar
m\sum_{\vec{k}}\gamma(\vec{k})\left(\hat{f}^\dagger(\vec{k})\hat{f}(\vec{k})+
\hat{g}^\dagger(\vec{k})\hat{g}(\vec{k})+1\right)
\quad .
\]
\end{itemize}
The vacuum energy is the sum of five contributions: if
$\bigtriangleup=\sum_{j=1}^2 \, \frac{\partial}{\partial x_j}\cdot
\frac{\partial}{\partial x_j}$ denotes the Laplacian,
\[
\Delta E^{(1)}_0=\sum_{\vec{k}}\sum_\alpha {\hbar m\over 2}
\omega(\vec{k})=\frac{3\hbar m}{2}{\rm
Tr}[-\bigtriangleup+1]^{{1\over 2}} \, \, , \, \, \Delta
E^{(2)}_0=\sum_{\vec{k}}{\hbar m\over 2} \nu(\vec{k})=\frac{\hbar
m}{2}{\rm Tr}[-\bigtriangleup+\kappa^2]^{{1\over 2}}
\]
\[
\Delta E^{(3)}_0=\sum_{\vec{k}}{\hbar m\over 2}
\omega(\vec{k})=\frac{\hbar m}{2}{\rm
Tr}[-\bigtriangleup+1]^{{1\over 2}}\quad , \quad \Delta
E^{(4)}_0=\sum_{\vec{k}}\hbar m \gamma(\vec{k})=\hbar m{\rm
Tr}[-\bigtriangleup]^{{1\over 2}}
\]
\[
E^{(5)}_0=-\sum_{\vec{k}}\hbar m \omega(\vec{k})=-\hbar m{\rm
Tr}[-\bigtriangleup+1]^{{1\over 2}}
\]
come from the vacuum fluctuations of the vector boson, Higgs, real
Goldstone, complex Goldstone, and ghost fields. Ghost fluctuations,
however, cancel the contribution of temporal vector bosons and real
Golstone particles, and the vacuum energy in the planar semi-local
AHM is due only to Higgs particles, complex Goldstone bosons, and
transverse massive vector bosons:
\[
\Delta E_0=\sum_{r=1}^5 \Delta E^{(r)}_0=\hbar m{\rm
Tr}[-\bigtriangleup+1]^{{1\over 2}}+{\hbar m\over 2}{\rm
Tr}[-\bigtriangleup+\kappa^2]^{{1\over 2}}+\hbar m{\rm
Tr}[-\bigtriangleup]^{{1\over 2}}  \qquad .
\]

\subsection{{\bf Semi-local self-dual topological solitons}}

At the critical point between Type I and Type II superconductivity,
$\kappa^2=1$, the energy can be arranged in a Bogomolny splitting:
\[
E=\frac{m^2}{2 e^2} \int d^2 x \left( ||D_1 \Phi \pm i D_2 \Phi||^2
+ [ F_{12} \pm {\textstyle\frac{1}{2}} (\Phi^\dagger \Phi-1) ]^2
\right)+\frac{m^2}{2}\frac{|g|}{e^2} \quad , \quad g= \int d^2 x
F_{12}=2{\pi l} \, \, , \, \, l\in{\mathbb Z} \quad .
\]
Therefore, the solutions of the first-order equations $D_1 \Phi \pm
i D_2 \Phi=0=F_{12} \pm \frac{1}{2} (\Phi^\dagger\Phi-1)$ are
absolute minima of the energy, hence stable, in each topological
sector with a classical mass proportional to the magnetic flux. It
has been shown in \cite{GORS} that there is a $4l$-dimensional
moduli space of such solitonic solutions interpolating between the
Nielsen-Olesen -NO in the sequel- vortices of the Abelian Higgs
model and the ${\mathbb CP}^1$-lumps of the planar non-linear sigma
model.

Assuming a purely vorticial vector field plus the spherically
symmetric ansatz
\begin{eqnarray*}
\phi_1(x_1,x_2) = f(r) {\rm cos}l\theta \quad &,& \quad
\phi_2(x_1,x_2) = f(r) {\rm sin}l\theta \\ \phi_3(x_1,x_2) = h(r)
{\rm cos}(\lambda+n\theta) \quad &,& \quad \phi_4(x_1,x_2) = h(r)
{\rm sin}(\lambda+n\theta)  \qquad , \qquad \lambda\in{\mathbb C} \,
, \, n\in {\mathbb Z}
\\ A_1(x_1,x_2) =-l \frac{\alpha(r)}{r}{\rm sin}\theta \quad &,&
\quad A_2(x_1,x_2) = l \frac{\alpha(r)}{r}{\rm cos}\theta \quad ,
\end{eqnarray*}
$g= - \oint_{r=\infty} dx_i A_i = -l\oint_{r=\infty}{
[x_2dx_1-x_1dx_2]\over r^2}=2 \pi l$, the first-order equations
reduce to{\footnote{The upper (lower) signs correspond to l and n
positive (negative). Finite energy solutions only exists if
$|n|<|l|$.}}
\[
{1\over r} {d \alpha \over d r}(r)= \mp \frac{1}{2 l}
(f^2(r)+h^2(r)-1) \quad , \quad {d f\over d r}(r) = \pm \frac{l}{r}
f(r)[1-\alpha(r)] \quad , \quad {d h\over d r}(r) = \pm \frac{l}{r}
h(r)[\frac{n}{l}-\alpha(r)]\qquad ,
\]
to be solved together with the boundary conditions
\begin{eqnarray}
&&{\displaystyle \lim_{r\rightarrow\infty}} f(r) = 1 \hspace{1cm} ,
\hspace{1cm} {\displaystyle \lim_{r\rightarrow\infty}}h(r) = 0
\hspace{1cm},\hspace{1cm} {\displaystyle
\lim_{r\rightarrow\infty}}\alpha(r) = 1 \nonumber \\ &&f(0) =0
 \hspace{1.5cm}
, \hspace{1.3cm}h(0)=h_0
\delta_{n,0}\hspace{1.2cm},\hspace{1.2cm}\alpha(0)=0 \,\,
\label{eq:sembc} ,
\end{eqnarray}
required by energy finiteness plus regularity at the origin (center
of the vortex). A partly numerical, partly analytical procedure
explained in detail in \cite{AJmW3} provides the field profiles
$f(r)$, $\alpha(r)$ as well as the magnetic field $B(r)={l\over
2r}\frac{d\alpha}{dr}$ and the energy density:
\[
 {\cal E}(r)={1\over 8}({1\over
l^2}+1)(1-f^2(r)-h(r)^2)^2+{l^2f^2(r)\over
r^2}(1-\alpha(r))^2+\frac{l^2 h(r)^2}{r^2}({|n|\over
|l|}-\alpha(r))^2 \qquad .
\]

We have worked completely in the last Reference the $l=1$,
$n=0,\lambda=0$ case and plotted the field profiles and the energy
density for four values of $h_0$. The physical meaning of the
parameter $h_0$, giving the size and the phase of the $\Phi_2$ field
for the solution at the origin, is also explained there. We remark
that solutions with $h_0=0$ are the NO vortices embedded in this
system and the growth of $h_0$ corresponds to the spread of the
energy density of the generic SSTS solutions. Solutions
with$|h_0|=1$ are the $CP^1$-lumps with energy density homogeneously
distributed over the whole plane.

\subsection{{\bf Casimir energy of semi-local self-dual topological solitons}}

Let us consider small fluctuations around vortices $
\Phi(x_0,\vec{x})= S(\vec{x})+\delta S(x_0,\vec{x}) \hspace{0.3cm} ,
\hspace{0.3cm} A_k(x_0,\vec{x})=V_k(\vec{x})+\delta a_k
(x_0,\vec{x})$, where by $S(\vec{x})$ and $V_k(\vec{x})$ we
respectively denote the scalar and vector field of the semi-local
vortex solutions. Working in the Weyl/background gauge
\[
A_0(x_0,\vec{x})=0 \qquad \qquad , \qquad \qquad \partial_j\delta
a_j(x_0,\vec{x})+ \frac{i}{2}(S^\dagger(\vec{x}) \delta
S(x_0,\vec{x})-\delta S^\dagger(x_0,\vec{x}) S(\vec{x}))=0  \quad ,
\]
the classical energy up to ${\cal O}(\delta^2)$ order is:
\[
H^{(2)}+H^{(2)}_{{\rm g.f.}}+H^{(2)}_{{\rm ghost}}={v^2\over 2}\int
\, d^2x\left\{\frac{\partial\delta\xi^T}{\partial
x_0}\frac{\partial\delta\xi}{\partial x_0}+\delta
\xi^T(x_0,\vec{x})K\delta\xi(x_0,\vec{x})+\delta\chi^*(\vec{x})K^G\delta\chi(\vec{x})\right\}
\quad ,
\]
where
\[
\delta\xi(x_0,\vec{x})=\left(\begin{array}{c} \delta a_1 (x_0,\vec{x}) \\
\delta a_2 (x_0,\vec{x}) \\ \delta S_1^1(x_0,\vec{x}) \\ \delta
S_1^2(x_0,\vec{x})\\ \delta S_2^1(x_0,\vec{x})\\ \delta
S_2^2(x_0,\vec{x})\end{array}\right) \qquad , \qquad
K^G=-\bigtriangleup +|S_1(\vec{x})|^2+|S_2(\vec{x})|^2  \qquad ,
\]
and
\begin{displaymath}
K=\left(\begin{array}{cccccc}
A&0&-2\nabla_1 S_1^2 &2\nabla_1 S_1^1 & -2\nabla_1 S_2^2 &2\nabla_1 S_2^1  \\
0&A&-2\nabla_2 S_1^2 & 2\nabla_2 S_1^1 & -2\nabla_2 S_2^2 & 2\nabla_2 S_2^1 \\
-2\nabla_1 S_1^2 &-2\nabla_2 S_1^2 &B&-2 V_k\partial_k&S_1^1S_2^1+S_1^2S_2^2 &S_1^1S_2^2-S_1^2S_2^1 \\
2\nabla_1 S_1^1 & 2\nabla_2 S_1^1&2 V_k \partial_k&B&-S_1^1S_2^2+S_1^2S_2^1 & S_1^1S_2^1+S_1^2S_2^2 \\
-2\nabla_1 S_2^2 &-2\nabla_2 S_2^2  &S_1^1S_2^1+S_1^2S_2^2  & -S_1^1S_2^2+S_1^2S_2^1 &C&-2 V_k\partial_k\\
2\nabla_1 S_2^1 & 2\nabla_2 S_2^1 &S_1^1S_2^2-S_1^2S_2^1  &
S_1^1S_2^1+S_1^2S_2^2 &2 V_k\partial_k&C
\end{array}\right) \qquad , \label{eq:bsfats1}\nonumber
\end{displaymath}
\begin{eqnarray*}
A&=&-\partial_k \partial_k+|S_1|^2+|S_2|^2 \qquad , \qquad
B=-\partial_k \partial_k+\frac{1}{2}(3|S_1|^2+|S_2|^2+2V_k V_k-1) \quad ,\\
C&=&-\partial_k\partial_k+\frac{1}{2}(|S_1|^2+3|S_2|^2+2V_k
V_k-1)\,\, \quad  , \quad \nabla_j S_M^a=\partial_j
S_M^a+\varepsilon^{ab} V_jS_M^b \qquad .
\end{eqnarray*}

The general solutions of the linearized field equations
\[
\frac{\partial^2\delta\xi_A}{\partial
x_0^2}(x_0,\vec{x})+\sum_{B=1}^6 \,
K_{AB}\cdot\delta\xi_B(x_0,\vec{x})=0 \hspace{2cm} , \hspace{2cm}
K^G \delta\chi(\vec{x})=\left(-\bigtriangleup
+|s(\vec{x})|^2\right)\delta\chi(\vec{x})=0
\]
are the eigenfunction expansions (the prime means that zero modes
are not included)
\begin{eqnarray*}
\delta\xi_A^\prime (x_0,\vec{x})&=&{e\over
mL}\sqrt{\frac{\hbar}{m}}\cdot
\sum_{\vec{k}}\sum_{I=1}^4\frac{1}{\sqrt{2\omega(\vec{k})}}\left[a^*_I(\vec{k})e^{i\omega(\vec{k})
x_0}u^{(I)*}_A(\vec{x};\vec{k})+a_I(\vec{k})e^{-i\omega(\vec{k})
x_0}u_A^{(I)}(\vec{x};\vec{k})\right] \\ &+& {e\over
mL}\sqrt{\frac{\hbar}{m}}\cdot
\sum_{\vec{k}}\sum_{I=5}^6\frac{1}{\sqrt{2\gamma(\vec{k})}}\left[a^*_I(\vec{k})e^{i\gamma(\vec{k})
x_0}u^{(I)*}_A(\vec{x};\vec{k})+a_I(\vec{k})e^{-i\gamma(\vec{k})
x_0}u_A^{(I)}(\vec{x};\vec{k})\right]
\end{eqnarray*}
\[
\delta\chi^\prime (x_0,\vec{x})={e\over
mL}\sqrt{\frac{\hbar}{m}}\cdot
\sum_{\vec{k}}\frac{1}{\sqrt{2\omega(\vec{k})}}\left[c(\vec{k})u^*(\vec{x};\vec{k})+d^*(\vec{k})u(\vec{x};\vec{k})\right]
\qquad ,
\]
where $A=1,2,3,4,5,6$ and by $u^{(I)}(k)$, $u(k)$ the non-zero
eigenfunctions of $K$ and $K^G$ are denoted respectively:
$I=1,2,3,4$, $Ku^{(I)}(\vec{x})=\omega(\vec{k})u^{(I)}(\vec{x})$,
$I=5,6$, $Ku^{(I)}(\vec{x})=\gamma(\vec{k})u^{(I)}(\vec{x})$,
$K^Gu(\vec{x})=\omega(\vec{k})u(\vec{x})$. Canonical quantization
\[
[\hat{a}_I(\vec{k}),\hat{a}_J^\dagger(\vec{q})]=\delta_{IJ}\delta_{\vec{k}\,\vec{q}}\quad
, \quad
\{\hat{c}(\vec{k}),\hat{c}^\dagger(\vec{q})\}=\delta_{\vec{k}\,\vec{q}}
\quad , \quad
\{\hat{d}(\vec{k}),\hat{d}^\dagger(\vec{q})\}=\delta_{\vec{k}\,\vec{q}}
\]
leads to the quantum free Hamiltonian
\begin{eqnarray*}
\hat{H}^{(2)}+\hat{H}^{(2)}_{{\rm g.f.}}+\hat{H}^{(2)}_{{\rm
Ghost}}&=&\hbar m \cdot\sum_{\vec{k}}\left[
\sum_{I=1}^4\,\omega(\vec{k})\left(\hat{a}_I^\dagger(\vec{k})\hat{a}_I(\vec{k})+{1\over
2}\right)+\sum_{I=5}^6\,\gamma(\vec{k})\left(\hat{a}_I^\dagger(\vec{k})\hat{a}_I(\vec{k})+{1\over
2}\right)\right] \\ &+& \hbar m \cdot\sum_{\vec{k}}\left[{1\over
2}\omega(\vec{k})\left(\hat{c}^\dagger(\vec{k})\hat{c}(\vec{k})+
\hat{d}^\dagger(\vec{k})\hat{d}(\vec{k})-1\right)\right]\quad ,
\end{eqnarray*}
and the ground state energy (all the modes non-occupied) of the
topological solitons reads:
\[
\bigtriangleup E_{\rm TS}=\frac{\hbar m}{2}{\rm STr}^*\, K^{{1\over
2}}=\frac{\hbar m}{2}{\rm Tr}^*\, K^{{1\over 2}}-\frac{\hbar
m}{2}{\rm Tr}^*\, (K^{{\rm G}})^{{1\over 2}}  \qquad ,
\]
where the star means that zero eigenvalues are not accounted for.
Note that the ghost fields are static in this combined
Weyl-background gauge and their vacuum energy is one-half with
respect to the time-dependent case. Only the Goldstone fluctuations
around the vortices must be subtracted. The zero-point vacuum energy
renormalization provides the Casimir energy for self-dual
($\kappa^2=1$) semi-local topological solitons:
\begin{equation}
\bigtriangleup M_{\rm TS}^C=\bigtriangleup E_{\rm TS}-\bigtriangleup
E_0={\hbar m\over 2}\left[{\rm STr}^*\, K^{{1\over 2}}-{\rm STr}\,
K_0^{{1\over 2}}\right] \label{eq:casv} \qquad .
\end{equation}

\subsection{{\bf Mass renormalization energy}}

In (2+1)-dimensional model only graphs with one or two external
lines are divergent in the vacuum Sector. We choose the following
counter-terms to cancel these divergences:
\begin{eqnarray*}
{\cal L}_{c.t.}^S &=& {\hbar\over 2}\left[2(\kappa^2+1)\cdot\,
I(1)+\kappa^2\cdot\, I(0)\right]\cdot\,
\left[\Phi^*_1(x^\mu)\Phi_1(x^\mu)+\Phi^*_2(x^\mu)\Phi_2(x^\mu)-1
\right]\\{\cal L}_{c.t.}^A &=&-\hbar [I(1)+I(0)]\cdot\,
A_\mu(x^\mu)A^\mu(x^\mu) \qquad , \qquad I(c^2)=\int \,
\frac{d^3k}{(2\pi)^3} \cdot \frac{i}{k^2-c^2+i\varepsilon} \qquad .
\end{eqnarray*}
Therefore,
\[
S_{c.t.}=\frac{\hbar}{2}\int\, d^3x \,
\left\{\left[2(\kappa^2+1)\cdot\, I(1)+\kappa^2\cdot\,
I(0)\right]\cdot\,
\left[2H+H^2+G^2+2|\varphi|^2\right]-2\left[I(1)+I(0)\right]\cdot
A_\mu A^\mu\right\}
\]
must be added to the bare action (\ref{eq:act}) to tame the
divergences arising in one-loop order. This specific choice fixes
finite renormalizations according to the following criteria:
\begin{enumerate}

\item We have used a minimal subtraction scheme taking care only of
infinite quantities.

\item By doing this, the choice of scalar field counter-terms sets
the no-tadpole condition for the critical value $\kappa^2=1$ between
Type I and Type II superconductivity, precisely the regime in which
we are interested. Vanishing of the tadpole ensures no modification
of the VEV $<\Phi> = (1 , 0)^T$ at one-loop level. This condition is
standard in the computation of one-loop mass shifts to
supersymmetric and non-supersymmetric kinks and vortices, see
\cite{BRvNV} and \cite{AJMW1}.

\item Considering no finite counter-terms for the derivative terms
of  the Higgs, $H$, and Goldstone, $G$, $\varphi$, fields, as well
as their three-valent and four-valent vertices, sets the poles of
their masses at their tree levels: $m_H=\kappa$, $m_G=1$,
$m_\varphi=1$, with residue one.

\item The mass counter-term for the vector boson field plus the no
addition of finite counter-terms for derivatives and three- and
four-valents vertices of this field keeps also the vector boson mass
at its tree level: $m_A=1$. Note that a mass term for the $A_\mu$
arises already at the tree level in the action (\ref{eq:act}) as a
consequence of the Higgs mechanism in the renormalizable gauge. This
point is crucial for staying at the critical value $\kappa^2=1$ in
the one-loop level.

\item When the zeta function regularization method is used in the
computation of one-loop mass shifts to non SUSY and SUSY kinks, the
large mass and heat kernel subtraction schemes are known to be
equivalent to the vanishing tadpole condition, see \cite{AJMW1},
\cite{BRvNV}, and \cite{BEKL}. Essentially this means that the
no-tadpole condition determines a contribution of the counter-terms
to the one-loop kink Hamiltonian energy density which exactly
cancels the contribution of the first coefficient of the
high-temperature heat function expansion $c_1(K)$ to the kink
Casimir energy. On the other hand, the contribution to the kink
Casimir energy of the zero-order coefficient is exactly canceled by
the zero-point vacuum energy renormalization. These two cancelations
together ensure that there are no divergences and no quantum
corrections in the energy in the infinite mass limit, as it should
be: there are no quantum fluctuations of infinite mass.

In the $(2+1)$-dimensional Abelian Higgs model also, only the
contributions of $c_0(K)$ and $c_1(K)$ to the vortex Casimir
energies would be non-zero (in fact, infinite) in the infinite mass
limit. The contribution of $c_0(K)$ is canceled like in the kink
case by subtracting the zero-point vacuum energy. The vanishing
tadpole condition, however, is necessary but not sufficient to
cancel the contribution of $c_1(K)$: one needs also the counter-term
to the vector boson mass considered above, see \cite{AJmW1}.

\item Finally, it would be possible to express all the divergent
Feynman amplitudes, up to finite parts, in terms, e.g., of the
divergent integral $I(1)$. Our choice of counter-terms, however,
respect the global $SU(2)$ symmetry which allows the existence a
priori of other topological solitons than the NO vortices.

A detailed calculation of some Feynman amplitudes needed to perform
this one-loop renormalization is offered in the last Appendix of
Reference \cite{AJMmW}.

\end{enumerate}

The contribution of these counter-terms to the one-loop mass shift
of the SSTS reads:
\[
\Delta M_{\rm TS}^R=\frac{\hbar\, m}{2}\,\int \, d^2x\,\left\{I(1)[
\,4(1-|S_1|^2-|S_2|^2)-2V_kV_k]+I(0)[(1-|S_1|^2-|S_2|^2)-2V_kV_k]\right\}
\]
and, formally, the total one-loop mass shift is: $\bigtriangleup
M_{\rm TS}=\bigtriangleup M_{\rm TS}^C+\bigtriangleup M_{\rm TS}^R$.

\section{{\bf The high-temperature one-loop vortex mass shift
formula}}

From the high-temperature expansion of the heat kernels
\[
{\rm Tr}e^{-\beta K} =  {e^{-\beta}\over 4\pi\beta}\cdot
\sum_{n=0}^\infty \,\sum_{A=1}^6 \,\beta^n [c_n]_{AA}(K)
\hspace{1.5cm} , \hspace{1.5cm} {\rm Tr}e^{-\beta K^{\rm G}} =
{e^{-\beta}\over 4\pi\beta}\cdot \sum_{n=0}^\infty\beta^n c_n(K^{\rm
G})
\]
the SSTS generalized zeta functions can be written in the form:
\[
\zeta_{K}(s)=\sum_{n=0}^\infty \left\{\sum_{A=1}^4
\frac{[c_n]_{AA}(K)}{4\pi\Gamma(s)}\cdot\gamma[s+n-1,1]+
\sum_{A=5}^6\frac{[c_n(K)]_{AA}}{4\pi
\Gamma(s)}\frac{1}{s+n-1}\right\}+{1\over\Gamma(s)}\int_1^\infty
{\rm Tr}^* e^{-\beta K} \, d\beta
\]
\[
\zeta_{K^G}(s)=\sum_{n=0}^\infty
c_n(K^G)\cdot\frac{\gamma[s+n-1,1]}{4\pi\Gamma(s)}+{1\over\Gamma(s)}\int_1^\infty
\, d\beta \, {\rm Tr}^* e^{-\beta K^{G}} \qquad .
\]
The diagonal Seeley coefficients $[c_n]_{AA}(K)$ of the K-heat
function high-T expansion (resp. the Seeley coefficients $c_n(K^G)$)
are the integrals over the whole plane of the Seeley densities
$[c_n]_{AA}(\vec{x},\vec{x};K)$ which arise in the associated K-heat
kernel expansion (resp. the Seeley densities
$c_n(\vec{x},\vec{x};K^G)$):
\[
[c_n]_{AA}(K)=\int \, d^2x \,[c_n]_{AA}(\vec{x},\vec{x};K) \qquad ,
\qquad c_n(K^G)=\int \, d^2x \,c_n(\vec{x},\vec{x};K^G) \qquad .
\]
Neglecting the entire part and setting a large but finite $N_0$, the
SSTS Casimir energies are regularized as
\begin{eqnarray*}
\Delta M_{\rm TS}^C(s)&=&\frac{\hbar\mu}{2}\left({\mu^2\over
m^2}\right)^s\left\{-\frac{4l}{\Gamma(s)}\int_0^1 d\beta \beta^{s-1}
+\sum_{n=1}^{N_0}\left[\sum_{A=1}^4 \,
[c_n]_{AA}(K)-c_n(K^G)\right]\cdot
\frac{\gamma[s+n-1,1]}{4\pi\Gamma(s)}+ \right.
\\ &+&\left.\sum_{A=5}^6\frac{[c_n(K)]_{AA}}{4\pi
\Gamma(s)}\frac{1}{s+n-1}\right\} \qquad ,
\end{eqnarray*}
where the $4l$ zero modes have been subtracted: the zero-point
vacuum renormalization amounts to ruling out the contribution of the
$c_0(K)$ and $c_0(K^G)$ coefficients. Also, $\Delta M_{\rm TS}^R$ is
regularized in a similar way
\[
\Delta M_{\rm TS}^R(s) = {\hbar\over 2\mu L^2}\left({\mu^2\over
m^2}\right)^s \left\{\zeta_{-\bigtriangleup+1} (s)\cdot \Sigma^{(1)}
(s(\vec{x}),V_k(\vec{x})) + \zeta_{-\bigtriangleup} (s)\cdot
\Sigma^{(0)} (s(\vec{x}),V_k(\vec{x})) \right\}
\]
\[
\Sigma^{(1)}(S,V_k)=4\int \, d^2x \, (1-|S_1|^2-|S_2|^2-{1\over
2}V_kV_k ) \qquad , \qquad  \Sigma^{(0)}(S,V_k)= \int \, d^2x
\,\left(1-|S_1|^2-|S_2|^2-2V_kV_k\right) \,\,\,.
\]
The physical limits $s=-{1\over 2}$ for $\Delta M_{\rm TS}^C$ and
$s={1\over 2}$ for $\Delta M_{\rm TS}^R$ are regular points of the
zeta functions. The contribution of the first coefficient of the
asymptotic expansion is not compensated by the contribution of the
mass renormalization counter-terms:
\begin{eqnarray*}
\Delta M_{TS}^{(1)C} (-1/2) &=& - {\hbar m \over 16 \pi }\left\{
\left(\Sigma^{(1)} (S,V_k)+2\int \, d^2x \,
\left|S_2\right|^2(x_1,x_2) \right) \cdot {\gamma[-1/2,1] \over
\Gamma(1/2)}\right. \\ &-&\left.\left(\Sigma^{(0)} (S,V_k)-2\int \,
d^2x \, \left|S_2\right|^2(x_1,x_2)\right)\cdot
\frac{2}{\Gamma(1/2)} \right\} \qquad ,
\end{eqnarray*}
\[
\Delta M_{TS}^R (1/2)  =  {\hbar m \over 16 \pi} \cdot\left\{
\Sigma^{(1)} (S,V_k) \cdot
 {\gamma[-1/2,1]\over
 \Gamma(1/2)}-{2\over\Gamma(1/2)}\cdot\Sigma^{(0)}(S,V_k)\right\}\qquad
 .
\]
Massless particles spoil the large mass subtraction criterion, see
\cite{BRvNV}, and we finally obtain the high-temperature one-loop
SSTS mass shift formula:
\begin{eqnarray}
\Delta M_{TS}&=& -{\hbar m \over 16\pi\sqrt{\pi}} \left[
\sum_{n=2}^{N_0}\,\left\{ [\sum_{A=1}^4[c_n(K)]_{AA}-c_n(K^{\rm G})]
\cdot\gamma[n-\frac{3}{2},1]+\sum_{A=5}^6\frac{[c_n(K)]_{AA}}{n-\frac{3}{2}}\right\}+4
l\cdot 8\pi \right] \nonumber\\&-&\frac{\hbar
m}{8\pi\sqrt{\pi}}\cdot \int \, d^2x \,
\left|S_2\right|^2(x_1,x_2)\cdot \left(\gamma[-{1\over
2},1]-2\right)\qquad \qquad . \label{eq:olslms}
\end{eqnarray}
\section{Numerical results}
Numerical methods are now implemented in a two-step procedure.
First, the Seeley densities are found by means of a symbolic program
run in a Mathematica environment on a PC. Second, numerical
integration of the Seeley densities on a disk of (non-dimensional)
radius $R=10^4$ allows us to compute the heat kernel coefficients.
We thus find, by setting $N_0=6$ and $l=1$, the following numerical
results for one-loop mass shifts of semi-local self-dual topological
solitons
\begin{eqnarray*}
M_{\rm TS}^{l=1}(h_0=0.1)&=&m\left(\frac{\pi v}{e}-1.55133
\hbar\right)+o(\hbar^2)\hspace{0.3cm} , \hspace{0.3cm} M_{\rm
TS}^{l=1}(h_0=0.3)= m\left(\frac{\pi v}{e}-0.252586
\hbar\right)+o(\hbar^2)\\
M_{\rm TS}^{l=1}(h_0=0.6)&=& m\left(\frac{\pi v}{e}+6.41655
\hbar\right)+o(\hbar^2) \hspace{0.3cm} , \hspace{0.3cm} M_{\rm
TS}^{l=1}(h_0=0.9)=  m\left(\frac{\pi
v}{e}+60.9433\hbar\right)+o(\hbar^2)\, \, ,
\end{eqnarray*}
as compared with the one-loop mass of the embedded Nielsen-Olesen
vortex:
\[
M_{\rm TS}^{l=1}(h_0=0.0)=m\left(\frac{\pi v}{e}-1.67989
\hbar\right)+o(\hbar^2) \qquad \qquad .
\]
Our numerical results suggest a breaking of the classical
degeneracy, the NO vortices remaining as the ground states of the
topological sector with $l=1$. These results are reinforced by the
following qualitative argument. The long-distance  behavior of the
Seeley densities is:
\begin{enumerate}
\item Embedded ANO vortex $h_0=0.0$: $2\pi r{\rm
tr}c_1^I(r)\propto {1\over r}$, $2\pi r{\rm tr}c_1^O(r)\propto
{1\over r}$, $2\pi r{\rm tr}c_2^I(r)\simeq {\cal O}({1\over r^3})$,
$2\pi r{\rm tr}c_2^O(r)\simeq{\cal O}({1\over r^3})$, $2\pi r{\rm
tr}c_n^I(r)\simeq {\cal O}(e^{-c r})$, $2\pi r{\rm tr}c_n^O(r)\simeq
{\cal O}(e^{-c r})$, $n>2$, when $r\rightarrow\infty$.
\item Semi-local topological soliton $h_0>0.0$: $2\pi r{\rm tr}c_n^I(r)\propto
{1\over r}$, $2\pi r{\rm tr}c_n^O(r)\propto {1\over r}$, $\forall
n$, when $r\rightarrow\infty$.
\end{enumerate}
If $h_0=0$, only the $c_1$ coefficient diverges, like $\log R$, but
its contribution is cancelled by mass renormalization counter-terms.
If $h_0>0$, all the Seeley coefficients are logarithmically
divergent and infrared divergences grow out of control.

\end{document}